# Selective Filtering of Photonic Quantum Entanglement via Anti-Parity-Time Symmetry

Mahmoud A. Selim[1], Max Ehrhardt[2], Yuqiang Ding[3], Hediyeh M. Dinani[1], Qi Zhong[3,4,5], Armando Perez-Leija[3], Şahin K. Özdemir[4,5], Matthias Heinrich[2], Alexander Szameit[2], Demetrios N. Christodoulides[1,6*], and Mercedeh Khajavikhan[1,6*]

[1]Ming Hsieh Department of Electrical and Computer Engineering, University of Southern California, CA 90089, USA
[2]Institut für Physik, Universität Rostock, Albert-Einstein-Straße 23, 18059 Rostock, Germany
[3]CREOL, College of Optics and Photonics, University of Central Florida, Orlando, Florida 32816-2700, USA
[4]Department of Engineering Science and Mechanics, The Pennsylvania State University, University Park, Pennsylvania 16802, USA
[5]Department of Electrical and Computer Engineering, Saint Louis University, Saint Louis, Missouri 63103, USA
[6]Department of Physics and Astronomy, University of Southern California, Los Angeles, CA, 90089, USA

## Abstract

Entanglement is a key resource for quantum computing, sensing, and communication, however it is highly susceptible to decoherence. To address this, quantum optics has explored filtering techniques like photon ancillas and Rydberg atom blockade to restore entangled states. Here, we introduce a an entirely new approach to entanglement retrieval exploiting non-Hermitian systems. By employing an anti-parity-time two-state guiding configuration, we demonstrate efficient extraction of entanglement from any input state. This filter is implemented on a lossless waveguide network using Lanczos transformations, consistent with Wigner-Weisskopf theory. This scheme achieves near-unity fidelity under single- and two-photon excitation and is scalable to higher photon levels while remaining robust against decoherence during propagation. Our work offers new insights into using non-Hermitian symmetries to address central challenges in quantum technologies.

Entanglement is a fundamental aspect of quantum mechanics, representing a unique and powerful form of nonclassical correlations between particles, and has far-reaching implications for quantum technologies. In quantum communications, for instance, the ability to manipulate entangled photon states underpins secure quantum key distribution (*1*, *2*), while in quantum computing, entanglement serve as the basis for the inherent parallelism that exponentially enhances computational capabilities (*3–6*) Similarly, in quantum sensing, entangled photons provide increased sensitivity and noise resilience that exceed the classical limit (*7*). Yet, the intrinsic fragility associated with entanglement poses a challenge, whereby minimal environmental interactions can destroy the delicate quantum superposition leading to a collapse into mixed or classical states (*8*). While such deterioration is almost universally anticipated in the presence of loss, it remains unclear to what extent non-Hermiticity can preserve or even restore this resource in a manner that is both scalable and ultimately efficient.

To retrieve an entangled state that has decomposed into a mixed state, a targeted approach can be used to selectively eliminate its classical components. This strategy closely resembles those associated with classical optical filters designed to isolate specific degrees of freedom of light, such as wavelength or polarization (*9*). In quantum optics, various methodologies for entanglement filtering have been explored, including schemes utilizing photon ancillas (*10*, *11*) or the nonlinear response of Rydberg atoms (*12*). Given that filters are inherently non-Hermitian entities, an intriguing question arises: can dissipation be engineered within specific non-conservative configurations to effectively restore entanglement from a mixed input state (*13*, *14*)?

Non-Hermitian systems have been extensively investigated in classical optical contexts, revealing a host of counterintuitive phenomena (*15*), including phase transitions (*16*), topological chirality (*17–20*), unidirectional invisibility (*21*), laser mode management (*22*, *23*), loss-induced transparency (*24*, *25*), and enhanced sensitivity (*26–28*), among others. Here, we leverage the distinctive properties of photonic non-Hermitian anti-parity-time (APT) symmetric configurations (*29*, *30*) to realize a class of structures with functionalities in the quantum regime. Our approach isolates a desired entangled state within a bosonic subspace, thereby providing a highly versatile linear mechanism for state selection through photon-photon interference. Importantly, this configuration functions as a decoherence-free subspace (*31*), preserving quantum states against dephasing while enhancing the robustness of quantum information processing.

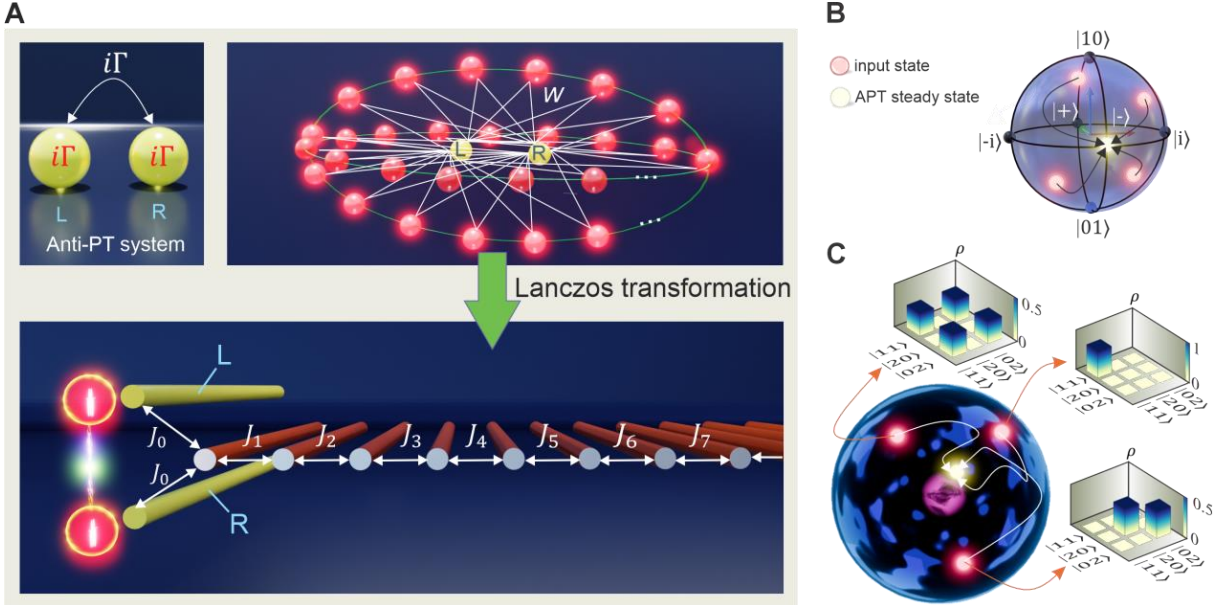

**Fig. 1. All-photonic scalable entanglement APT filter.** (**A**) Schematic of an APT symmetric configuration, representing two lossy cavities or waveguides with an imaginary coupling coefficient (upper left panel). Through Wigner-Weisskopf theory, dissipation is modeled through coupling to a continuum of elements with uniformly increasing detuning (upper right panel). Utilizing the Lanczos transformations, the system is reformulated into an array of coupled waveguides that selectively transmits the desired entangled state (lower panel). (**B**) APT dynamics on the single-photon Bloch sphere. Regardless of the input single-photon state, the output irreversibly yields ($|\psi^{(1)}\rangle = (|10\rangle - |01\rangle)/\sqrt{2}$). (**C**) Bloch ball for two-photon states. In this case, the output state consistently evolves into the two-photon entangled state $|\psi^{(2)}\rangle = (|20\rangle + |02\rangle)/2 - |11\rangle/\sqrt{2}$.

**Photonic entanglement filter system--theory**

Anti-parity-time symmetry, a subclass of non-Hermitian systems, is associated with Hamiltonians $\mathcal{H}$ that anti-commute with the parity-time ($\hat{\mathcal{P}}\hat{\mathcal{T}}$) operator, i.e., $\{\hat{\mathcal{P}}\hat{\mathcal{T}}, \mathcal{H}\} = \hat{\mathcal{P}}\hat{\mathcal{T}}\mathcal{H} + \mathcal{H}\hat{\mathcal{P}}\hat{\mathcal{T}} = 0$. In optics, APT symmetry can be established in scenarios where two elements (e.g. waveguides or cavities), labeled L and R, are dissipatively coupled (Fig. 1A upper left panel) with a Hamiltonian given by:

$$\mathcal{H}_{eff} = -i\Gamma(\sigma_x + I), \quad (1)$$

where $\sigma_x$ denotes the $x$ Pauli operator, $I$ is the identity matrix, and $\Gamma$ represents the coupling coefficient. Realizing an APT-symmetric system is challenging, as it requires the dissipative coupling $i\Gamma$ to precisely match the local dissipations at sites L and R. Serendipitously, this symmetry emerges naturally in a mirrored Wigner-Weisskopf configuration (*32*), (Fig. 1A upper right panel), where two isoenergetic states happens to exhibit APT symmetry, provided both elements are coupled with equal strength $w$ to a Hermitian bath consisting of a continuum of equidistantly spaced energy levels [Supplementary Text I (*33*)]. In this case, the effective Hamiltonian governing the dynamics at sites L and R is precisely given by the non-Hermitian Hamiltonian of Eq. (1)—a rather surprising result given that the infinite chain in between the two waveguides is itself conservative. Yet, even at this stage, achieving an optical realization of the configuration in Fig. 1A remains a formidable challenge. Geometrically, it is unfeasible to surround a cavity or waveguide site with large number of elements whose local eigenvalues continuously increase/decrease at a uniform rate. To circumvent this physical constraint, we use isospectral Lanczos transformations [Supplementary Text II (*33*)], which enable mapping the intermediate infinite chain onto a tridiagonal matrix with elements sharing identical local eigenvalues (*34*, *35*). This mapping holds as long as the exchange strength between nearest-neighbor elements is appropriately engineered (Fig. 1A lower panel) [Supplementary Text III (*33*)]. It should be emphasized that the collective quantum mechanics of a subsystem together with its environment is inherently Hermitian, and any presence of non-Hermiticity only emerges when projecting onto the subsystem via post-selection.

To benchmark our experimental and theoretical studies, we numerically model the Markovian dynamics of the reduced density matrix $\hat{\rho} = \sum_j p_j |\psi_j\rangle\langle\psi_j|$, where $p_j$ denotes the probability of the state $|\psi_j\rangle$, by using the standard Lindblad master equation (*36*):

$$\partial_z \rho(z) = -i(\mathcal{H}_{eff}\hat{\rho} - \hat{\rho}\mathcal{H}_{eff}^\dagger) + 2\Gamma \hat{a}_L \hat{\rho} \hat{a}_L^\dagger + 2\Gamma \hat{a}_R \hat{\rho} \hat{a}_R^\dagger = \mathcal{L}\rho. \qquad (2)$$

In general, the solution of the master equation ($\hat{\rho} = e^{\mathcal{L}z}\hat{\rho}(0)$) can be obtained from the system's eigenmodes after diagonalizing $\mathcal{L}$ (*37*). For instance, under one- and two-photon excitation conditions, one can show that the APT effective Hamiltonian of Eq. (1) allows a lossless eigenstate: (i) $|\psi^{(1)}\rangle = 1/\sqrt{2}(|10\rangle - |01\rangle)$ (also known as W-state) existing within the single-photon subspace, and (ii) $|\psi^{(2)}\rangle = 1/2(|20\rangle + |02\rangle) - |11\rangle/\sqrt{2}$ arising when the subspace engages two photons [Supplementary Texts IV, V, and VI (*33*)]. In contrast, the rest of the modes undergo loss, decaying over a propagation distance $z \gg 1/\Gamma$, as their corresponding eigenvalues exhibit a finite imaginary component. Note that these two lossless states $|\psi^{(1)}\rangle$ and $|\psi^{(2)}\rangle$ are path entangled. More broadly, any arbitrary $N$-photon excitation—whether on the Bloch sphere (for pure states) or within the Bloch ball (for mixed states)—irreversibly evolves towards a single point, corresponding to a specific entangled mode. This process is schematically illustrated in Fig. 1B for $N = 1$, where all states on the one-photon Bloch sphere converge after propagation to the W-state (*38*). Conversely, for $N = 2$, all possible excitations eventually yield $|\psi^{(2)}\rangle = 1/2(|20\rangle + |02\rangle) - |11\rangle/\sqrt{2}$ (Fig. 1C). Evidently, this behavior is consistent with that expected from an entanglement filter. In principle, the quantum dynamics in this altogether Hermitian setting (APT subsystem and Lanczos array) can be theoretically described by treating this waveguide array arrangement as a multi-port Hong-Ou-Mandel system [Supplementary Text IV (*33*)]. Moreover, this approach and that of Lindblad [Eq. (2)] yield identical results. We also note that in principle post-selecting the anti-coincidence events projects the $|\psi^{(2)}\rangle$ onto the two-photon NOON state which is a resource for quantum metrology [Supplementary Text V (*33*)].

**Experimental results**

To experimentally verify the filtering behavior of the APT system, we design a set of multi-elements guiding structures using Lanczos transformations. These arrangements are then fabricated via femtosecond direct laser writing in fused silica glass (*39*) [Supplementary Text VI (*33*)]. Samples with different lengths are prepared to enable the observation of the dynamics of the quantum system. In our experiments, the imaginary coupling factor is set to be $\Gamma = 0.25 \text{ cm}^{-1}$ and a Lanczos's array of 52 non-uniformly coupled waveguides is deployed. In all cases, the loss factor is experimentally characterized using a modified variable stripe technique [Supplementary Text III (*33*)]. As a first step, we monitor the classical evolution by measuring the intensity ratios at the output of the two waveguides using laser light at a wavelength of $\lambda = 810$ nm.

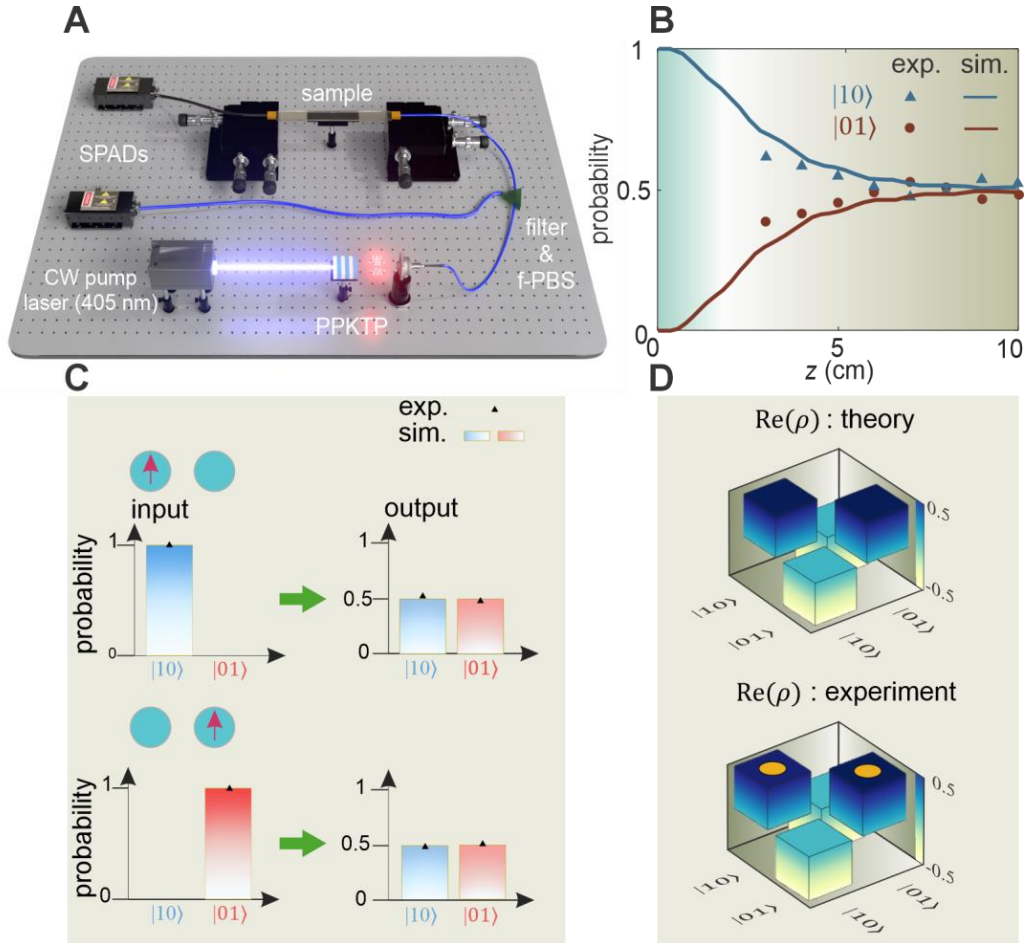

**Fig. 2. Single-photon response of the APT filter.** (**A**) Schematic of the experimental setup for single-photon measurements. Photon pairs are generated via type-II SPDC. One photon is routed into a fused silica chip containing laser-written photonic circuits through a single-mode polarization-maintaining fiber array, then collected using a multimode fiber. The other photon is directly connected to a single-photon detector to enable heralded measurements. Here, f-PBS: fiber polarizing beam splitter; BS: beam splitter; SPAD: single-photon avalanche diode. (**B**) Measured probability of finding a photon in each waveguide after varying propagation lengths. Solid lines represent theoretical predictions for the design parameter $\Gamma = 0.25\ cm^{-1}$. Measurement errors are within the symbol size. (**C**) Performance evaluation of the APT entanglement filter at $z = 10\ cm$, where single photons are injected via $|10\rangle$ and $|01\rangle$ states. (**D**) Comparison of the theoretical (top) and experimentally measured (bottom) density matrix $\rho = |\psi\rangle\langle\psi|$ at the APT filter output at $z = 10\ cm$. Experimental errors (orange cylinders) denote Poisson standard deviations, typically less than 2% in all cases.

The APT entanglement filter's response is then investigated under single-photon excitation by means of heralded detection. Entangled photon pairs are generated via type II spontaneous parametric down-conversion (SPDC) in a periodically poled potassium titanyl phosphate (PPKTP) crystal, converting a 405 nm pump photon into a pair of polarization-entangled signal and idler photons at 810 nm. These two photons are subsequently split using a fiber-coupled polarization beam splitter, after which one photon is used for heralding, while the other is coupled to one of the sites of the APT arrangement (Fig. 2A). The single-photon evolution dynamics are monitored within the basis states $|10\rangle$ or $|01\rangle$ as depicted in Fig. 2B, using structures of varying lengths (3 cm to 10 cm in 1 cm increments). Figure 2C shows the photon detection probability at the output of each waveguide element after a propagation of $z = 10$ cm, confirming that, irrespective of the input photon state $|10\rangle$ or $|01\rangle$, the system reaches an equilibrium with equal photon probabilities at both outputs ($P_{10} = P_{01} = 0.5$). However, the probability measurement by itself does not uniquely specify the quantum state. In general, this measurement only indicates that the output state is of the form $(|10\rangle + e^{i\phi}|01\rangle)/\sqrt{2}$, where $\phi$ can be an arbitrary phase. To characterize this state, we perform quantum state tomography using additional measurement configurations where we interfere the output of the two channels using a 3dB coupler after adding a $\pi/4$ phase shift in one of the arms. By considering these measurements, one can determine $\phi$ [Supplementary Texts VII and VIII (33)], which, as indicated by the density matrix in Fig. 2D, is equal to $\pi$ in this case. This observation corroborates that the output state is indeed $|\psi^{(1)}\rangle = (|10\rangle - |01\rangle)/\sqrt{2}$. Note that the orange cylinders in Fig. 2D highlight the experimental error.

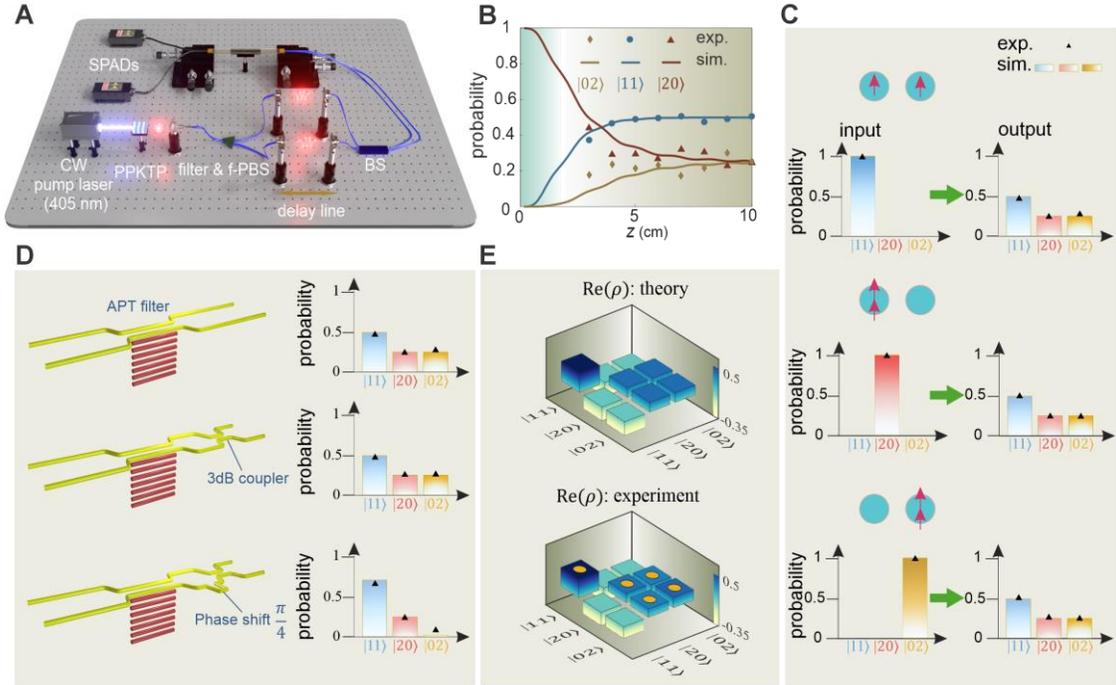

**Fig. 3. Two-photon dynamics of the APT filter.** (**A**) Experimental setup for examining the APT entanglement filter under two-photon excitation. Here, f-PBS: fiber polarizing beam splitter; BS: beam splitter; SPAD: Single-Photon Avalanche Diode. (**B**) The response of the APT filter for two-photon excitations, characterized by measuring samples with the same APT coupling parameter $\Gamma$ for different propagation distances. (**C**) Probabilities of detecting photons at the output of the APT filter under various two-photon excitation conditions. (**D**) Quantum state tomography measurements. Additional configurations based on a 50/50 coupler with phase shifts of 0 and $\pi/4$ in one of the arms are utilized to uniquely identify the phases. (**E**) Theoretically calculated (upper) and experimentally measured (lower) two-photon density matrix $\rho = |\psi\rangle\langle\psi|$ at $z = 10\ cm$. Experimental errors, indicated by orange cylinders, typically ranging between 1% and 4%.

We next consider the APT filtering action when two-photons are injected using the experimental setup depicted in Fig. 3A. In this case, timing of the photon pairs is achieved by leveraging the Hong–Ou–Mandel (HOM) interference effect (40) [Supplementary Text IX (33)]. After that, one arm of the HOM interferometer is directed towards the sample, while the other port is blocked. This arrangement and its variants allow us to excite the sample with basis vectors $|20\rangle$, $|02\rangle$, or $|11\rangle$. At the output, the sample is aligned with a two-element multimode fiber array having a pitch of 127 μm, to match the fan-out of the APT filter's waveguides. The multimode fibers are then routed to single-photon detectors, interfaced with a time tagger for Time-Correlated Single Photon Counting. As before, to characterize the evolution dynamics of the two-photon quantum state, we perform a series of measurements on samples with a constant imaginary coupling factor but varying lengths (3 cm to 10 cm). As anticipated, after sufficient propagation distance ($z \gg 1/\Gamma$), the output consistently converges to the system's attractor state, as shown in Fig. 3B for $|20\rangle$ input state. Next, we test the APT filter by launching the other aforementioned two-photon input states and monitoring their dynamics after propagation over a distance of $3 - 10$ cm. In all cases, the output state consistently exhibits the expected detection probabilities (Figs. 3C). These initial measurements suggest that the observed state is of the form $(|20\rangle + e^{j\phi_1}|02\rangle)/2 + e^{j\phi_2}|11\rangle/\sqrt{2}$. To identify the relative phases $\phi_1$, $\phi_2$ we next perform interferometric measurements to reconstruct the density matrix via quantum tomography techniques. By further examining the output state using a balanced and quarter-wave shifted 3dB coupler, we determine $\phi_1 \approx 0$ and $\phi_2 \approx \pi$ (see Fig. 3D). The observed entangled state is consistent with our theoretical predictions as shown in Fig. 3E.

The experiments above highlight the universality of APT filter across different photon subspaces. This is attributed to the presence of a solitary attractor in each subspace, facilitated by the interplay of non-Hermiticity and photon-photon interference. While our current setup limit us to $N = 2$ photons, the APT filter can function under $N$-photon excitation conditions, where the zero-loss quantum state attractor assumes the form [Supplementary Text X (33)]:

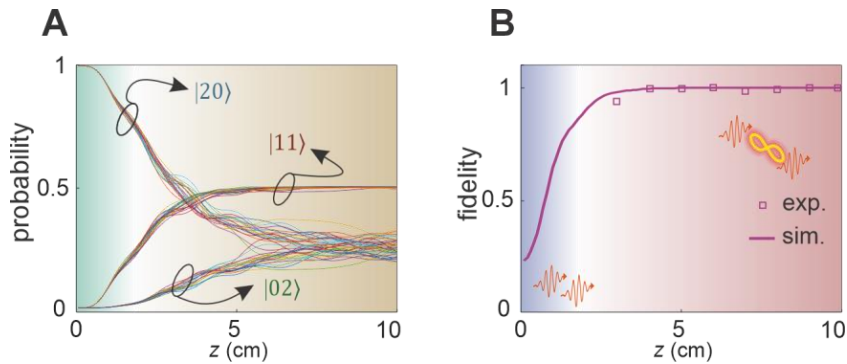

**Fig. 4. Resilience of APT entanglement filter.** **(A)** Ensemble response of the system's evolution with approximately 10% variations in the coupling coefficients. Even in the presence of this perturbation, the output still settles into the entangled state. **(B)** The evolution of the fidelity of APT state.

$$|\psi\rangle = \sum_{k=0}^{N}(-1)^k \sqrt{\frac{1}{2^N}\binom{N}{k}}|N-k,k\rangle. \tag{3}$$

Furthermore, these states are resilient to phase or structural perturbations manifested in the Lanczos array. In fact, when the eigenvector is perturbed after reaching a steady state, it automatically returns to equilibrium after propagating over a distance on the order of $z \gg 1/\Gamma$. This behavior is analogous to that of a decoherence-free subspace, where states propagate without being significantly affected by environmental noise (Fig. 4A) [Supplementary Text XI (*33*)]. In addition to its reduced implementation complexity, an advantage of the APT filter lies in its capacity to achieve the target state with high fidelity, circumventing the constraints associated with other filtering techniques. For example, measurement-based protocols typically exhibit lower fidelities, while double Rydberg excitation filters face intrinsic challenges in suppressing unwanted interactions. To quantify this aspect, we define the fidelity $\mathcal{F}$ through the diagonal elements of the density matrix using Bhattacharyya centroids (*41*), as $\mathcal{F} = \sum P_{exp} P_{meas}$, where, $P_{exp}$ represents the expected probability of the entangled wavefunction (e.g. $|\psi\rangle = (|20\rangle + |02\rangle)/2 - |11\rangle/\sqrt{2}$) for two-photon excitation), and $P_{meas}$ denotes the measured probability from the APT system. In this context, the measured fidelity of the APT entanglement filter exceeds 99% for $z > 1/\Gamma$ as shown in Fig. 4B. Finally, it should be noted that the filter introduced in this work has the capability to purify mixed states [Supplementary Text XII (*33*)].

**Concluding remarks**

We demonstrated that non-Hermiticity in the form of anti-parity-time symmetry can be harnessed to realize a class of entanglement filters. The APT structure was designed using a methodology based on isospectral Lanczos transformations and was experimentally validated under both single- and two-photon excitation conditions. The filtering response is achieved in a linear manner through the interplay of photon-photon interference and dissipation engineering. Furthermore, implementing APT systems within a completely Hermitian environment presents a promising path forward in non-Hermitian quantum mechanics, eliminating the need for absorbing or amplifying materials. Ultimately, by enabling the on-demand generation of entangled photons and nondestructive entanglement purification on-chip, this work sets the stage for advanced quantum technologies to be developed on integrated and compact platforms.

**Acknowledgements:** The authors gratefully acknowledge Prof. Yogesh Joglekar for valuable technical discussions.

**Funding:** This work was supported by the Air Force Office of Scientific Research (AFOSR) Multidisciplinary University Research Initiative (MURI) award on Programmable systems with non-Hermitian quantum dynamics (award no. FA9550-21-1-0202) (M.A.S., Y.D., H.M.D., Q.Z., A.P.L.,S.K.O., D.N.C., and M.K.), AFOSR MURI on Novel light-matter interactions in topologically non-trivial Weyl semimetal structures and systems (award no. FA9550-20-1-0322) (M.A.S., H.M.D., D.N.C., and M.K.), AFOSR MURI award ONR MURI award on the classical entanglement of light (award no. N00014-20-1-2789) (M.A.S., H.M.D., D.N.C., and M.K.), W.M. Keck Foundation (D.N.C.), MPS Simons collaboration (Simons grant no. 733682) (D.N.C.), and US Air Force Research Laboratory (FA86511820019) (D.N.C.). The Department of Energy (DESC0022282) (D.N.C., M.K. and M.A.S.). The authors thank C. Otto for preparing the high-quality fused silica samples used for the inscription of all photonic structures employed in this work. A.S. acknowledges funding from the Deutsche Forschungsgemeinschaft (grants SZ 276/9-2, SZ 276/19-1, SZ 276/20-1, SZ 276/21-1, SZ 276/27-1, and GRK 2676/1-2023 'Imaging of Quantum Systems', project no. 437567992). A.S. also acknowledges funding from the Krupp von Bohlen and Halbach Foundation as well as from the FET Open Grant EPIQUS (grant no. 899368) within the framework of the European H2020 programme for Excellent Science. A.S. and M.H. acknowledge funding from the Deutsche Forschungsgemeinschaft via SFB 1477 'Light–Matter Interactions at Interfaces' (project no. 441234705).
**Author Contributions** The idea was conceived by M.A.S., A.P.L., D.N.C., and M.K., whereas M.A.S., M.E., A.P.L., Q.Z. Y.D., M.H. designed the structures and fabricated the samples, M.A.S. and M.E. performed the experiments. All authors discussed the results and co-wrote the manuscript.
**Competing interests:** The authors declare no competing interests.
**Data and materials availability:** All data needed to replicate the work are present either in the main text or supplementary material. Raw data are available at Dryad online repositories (*42*).

**Supplementary Materials:**

Supplementary Text

Figs. S1 to S10

Table S1

References (43-45)

**Supplementary Text I. Anti-parity-time symmetry in engineered environments**
In the framework of open quantum systems, a configuration of interest interacting with an environment is often modelled as a set of harmonic oscillators (the system-of-interest) coupled to an infinite collection of non-interacting harmonic oscillators (the environment). In optics, such a system could be implemented using a single-mode waveguide (cavity) coupled to an infinite chain of uncoupled waveguides (cavities) as illustrated in Fig. S1. In this section we show that this problem can be solved using the Wigner-Weisskopf model in which a discrete level is coupled to a quasi-continuum of states (*32*).

The propagation dynamics of a single photon traversing the waveguide system (R) shown in Fig. S1A is governed by the following set of differential equations:

$$i\frac{d}{dz}c_R(z) = w \sum_{k=-\infty}^{+\infty} c_k(z) e^{-ikz}, \quad (S1)$$

$$i\frac{d}{dz}c_k(z) = w e^{ikz} c_R(z), \quad (S2)$$

here, $c_{R,k}(z)$ represents the probability amplitude of finding the photon at the system (environment) waveguide $R(k)$, $z$ is the propagation coordinate, while $w$ denote the coupling coefficient between the system waveguide and the environment waveguides. Direct integration of Eq. (S2) yields

$$c_k(z) = -iw \int_0^z e^{ikz'} c_R(z') dz' + c_k(0). \quad (S3)$$

Assuming that at $z = 0$ the environment waveguides are in the vacuum state, that is, $c_k(0) = 0$, then we obtain

$$c_k(z) = -iw \int_0^z c_R(z') e^{ikz'} dz'. \quad (S4)$$

Substituting Eq. (S4) into Eq. (S1) yields

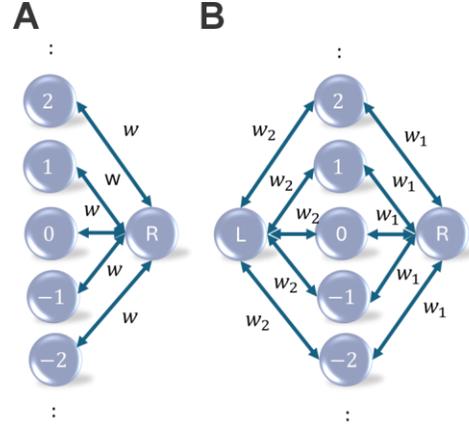

**Fig. S1. Schematic representation of qubit-environment coupling in open quantum systems.** (**A**) A single waveguide R coupled to an environmental bath modelled as an infinite waveguide array. (**B**) Two waveguides, labelled L and R, coupled to a common environment. Note that the cavities or the waveguides are not directly coupled to each other but via the shared environment they are both coupled to.

$$\frac{d}{dz}c_R(z) = -\frac{\Gamma}{\pi} \int_0^z c_R(z') \left[\sum_{k=-\infty}^{+\infty} e^{ik(z'-z)}\right] dz', \quad (S5)$$

where we have defined $w^2 = \frac{\Gamma}{\pi}$. Assuming a quasi-continuum of waveguides allows us to approximate $\sum_{k=-\infty}^{+\infty} e^{ik(z'-z)} \approx \int_{-\infty}^{\infty} e^{ik(z'-z)} dk = 2\pi\delta(z'-z)$, such that Eq. (S5) becomes:

$$\frac{d}{dz}c_R(z) = -2\Gamma \int_0^z c_R(z')\delta(z'-z)dz' = -\Gamma c_R(z). \quad (S6)$$

This demonstrate that the probability amplitude along the system waveguide decays exponentially according to:

$$c_R(z) = c_R(0)e^{-\Gamma z}. \quad (S7)$$

In the same manner, the single-photon dynamics of two uncoupled waveguides (R and L), jointly coupled to an environment, as illustrated in Fig. S1B, is described by the set of equations

$$i\frac{d}{dz}c_L(z) = w_2 \sum_{k=-\infty}^{+\infty} c_k(z)e^{-ikz},$$

$$i\frac{d}{dz}c_k(z) = \big(w_2 c_L(z) + w_1 c_R(z)\big)e^{ikz}, \qquad (S8)$$

$$i\frac{d}{dz}c_R(z) = w_1 \sum_{k=-\infty}^{+\infty} c_k(z)e^{-ikz}.$$

Following similar steps as in the previous case and assuming $w_1 = w_2 = w$, we can get $c_k(z) = -iw \int_0^z w e^{ikz'}\big(c_L(z') + c_R(z')\big)dz' + c_k(0)$. Then assuming that environment is vacuum (i.e., $c_k(0) = 0$) and substituting in first rate equation in Eqs. (S8), we can obtain $\frac{d}{dz}c_R(z) = -\frac{\Gamma}{\pi}\int_0^z \big(c_R(z') + c_L(z')\big)\big[\sum_{k=-\infty}^{+\infty} e^{ik(z'-z)}\big]dz' = \Gamma\big(c_R(z) + c_L(z)\big)$, and these equations can be simplified to:

$$i\frac{d}{dz}c_L(z) = -i\Gamma\big(c_L(z) + c_R(z)\big),$$
$$i\frac{d}{dz}c_R(z) = -i\Gamma\big(c_L(z) + c_R(z)\big). \qquad (S9)$$

The Hamiltonian for this system can be written as:

$$H = \begin{pmatrix} -i\Gamma & -i\Gamma \\ -i\Gamma & -i\Gamma \end{pmatrix} = -i\Gamma \begin{pmatrix} 1 & 1 \\ 1 & 1 \end{pmatrix}. \qquad (S10)$$

which corresponds to the APT symmetric Hamiltonian as described in Ref. (*30*). This matrix has two eigenvectors $\hat{s}_+ = \frac{1}{\sqrt{2}}\begin{pmatrix}1\\1\end{pmatrix}$ and $\hat{s}_- = \frac{1}{\sqrt{2}}\begin{pmatrix}1\\-1\end{pmatrix}$ with the corresponding eigenvalues $2i\Gamma$ and $0$, respectively. The first mode, (i.e., $\hat{s}_+$) decays exponentially as it propagates through the APT structure. In contrast, the second mode, $\hat{s}_-$, remains invariant and evolves without any loss. Similarly, if the input is a superposition of these two modes, the output will relax to the steady state of the APT system, represented by $\hat{s}_-$, exhibiting no dissipation afterward. As emphasized in the main text, this waveguide system cannot be physically implemented, as it requires the fabrication of an infinite array of uncoupled waveguides, each waveguide being simultaneously coupled to the waveguides conforming the system. In the next section, we demonstrate that these types of systems can be mapped onto two physically realizable waveguide systems by means of the so-called Lanczos transformations.

**Supplementary Text II. Lanczos transformations**
As discussed in Part I, the imaginary coupling necessary for APT symmetry can be achieved using a two-qubit system that is equally coupled to the same environment. Although this setup could theoretically be implemented with evanescently coupled waveguide arrays, it is practically infeasible to surround each waveguide with an infinite array of detuned waveguides. Moreover, the analysis in Text I requires that the environment remains uncoupled, a condition that is unattainable with an infinite set of waveguides. Additionally, equal coupling coefficients from the left and right waveguides to the environment are essential.

To address these challenges, we employ the Lanczos transformation (*34, 35*) to effectively simulate the environmental effects for both the left and right waveguides. By symmetrically coupling these waveguides to a common environment, we can achieve the desired APT symmetry. Generally, as the geometric dimensionality of a coupled lattice increases, the range of interactions also expands when the Hamiltonian is represented in the conventional 2D matrix format. To establish comparable excitation dynamics between a central site in a multi-dimensional structure and the first site of a one-dimensional lattice with nearest-neighbor coupling, we utilize a tailored Lanczos algorithm to transform the Hamiltonian into a desired tridiagonal configuration. For an $m \times m$ Hamiltonian $\mathbf{H}$, the algorithm involves the following iterative steps:

Starting with a unit vector $\mathbf{v_1}$ in an $m$-dimensional Hilbert space, we perform the following iterative steps (34, 35):
Let $\mathbf{p'_1} = \mathbf{H}\mathbf{v_1}$, compute $\epsilon_1 = (\mathbf{p'_1})^\dagger \mathbf{v_1} \equiv (\mathbf{H}\mathbf{v_1})^\dagger \mathbf{v_1}$ and set $\mathbf{p_1} = \mathbf{p'_1} - \epsilon_1 \mathbf{v_1}$
    For $i = 2,3,4 \dots, m$:
    a) Define $C_{i-1} = ||\mathbf{p_{i-1}}||$, here $||.||$ is the Euclidian norm.
    b) If $C_{i-1} \neq 0$, set $\mathbf{v_i} = \mathbf{p_{i-1}}/C_{i-1}$; However, if $C_{i-1} = 0$, select a unit vector $\mathbf{v_i}$ is orthogonal to $\mathbf{v_1}$ to $\mathbf{v_{i-1}}$.
    c) Compute $\mathbf{p'_i} = \mathbf{H}\mathbf{v_i}$
    d) Compute $\epsilon_i = (\mathbf{p'_i})^\dagger \mathbf{v_i} \equiv (\mathbf{H}\mathbf{v_i})^\dagger \mathbf{v_i}$.
    e) Assume $\mathbf{p_i} = \mathbf{p'_i} - \epsilon_i \mathbf{v_i} - C_{i-1}\mathbf{v_{i-1}}$.
Construct the matrix $\mathbf{V}$ whose columns are $\mathbf{v_1}, \mathbf{v_2}, \dots, \mathbf{v_m}$. The matrix $\mathbf{H}^{(3)}$ is then formed as:

$$\mathbf{H}^{(3)} = \begin{bmatrix} \epsilon_1 & C_1 & 0 & 0 & \dots & 0 \\ C_1 & \epsilon_2 & C_2 & 0 & \dots & 0 \\ 0 & C_2 & \epsilon_3 & C_3 & \dots & 0 \\ \vdots & \vdots & \vdots & \vdots & \ddots & \vdots \\ 0 & 0 & 0 & 0 & \dots & \epsilon_m \end{bmatrix}. \quad (S11)$$

In our study, we meticulously tailor the tridiagonalization process by selecting a specific $\mathbf{v_1} = |m_a\rangle$, where $|m_a\rangle$ denotes the anchor site within the multi-dimensional lattice. This choice is critical, as discussed in the main manuscript, to ensure that the first site of the one-dimensional lattice accurately replicates the dynamics of the anchor site in the multi-dimensional lattice.

**Supplementary Text III. Coupling coefficients for simulation and experiments**
As discussed in Supplementary Text I, emulating an effective environment is essential for the dynamics of both single and two-qubit systems. Following the application of the Lanczos transformation to the system described in Supplementary Text I, the resulting coupling coefficients are presented in Table S1. For the single-qubit case, a waveguide array can be realized as illustrated in Fig. S2A. It is important to highlight that the coupling coefficients in Table S1 can be derived using the Lanczos transformation with large number of sites (e.g., $N = 1000$), followed by truncating the array to 50 elements. Similarly, the APT system can be realized by symmetrically coupling two uncoupled waveguides to the same artificial environment, as depicted in Fig. S2B. Since the two qubits are uncoupled in Fig. S1B, we can apply the Lanczos transformation to the system illustrated in Fig. S1A and couple the resulting Lanczos array to two isolated waveguides (Fig. S2B). This approach will replicate the dynamics for the configuration depicted in Fig. S1B. Crucially, this procedure does not alter the dynamics, as the reflection takes a sufficiently long distance to influence the evolution. To validate these theoretical predictions, fused silica waveguide arrays were fabricated using the coupling parameters listed in Table 1. Additionally, multiple samples with identical coupling but varying propagation lengths were produced to capture the system's dynamics and to measure the effective loss. As anticipated, these samples exhibit exponential decay, closely mimicking the spontaneous emission process of a single-qubit system coupled to the environment, as shown in Fig. S2C.

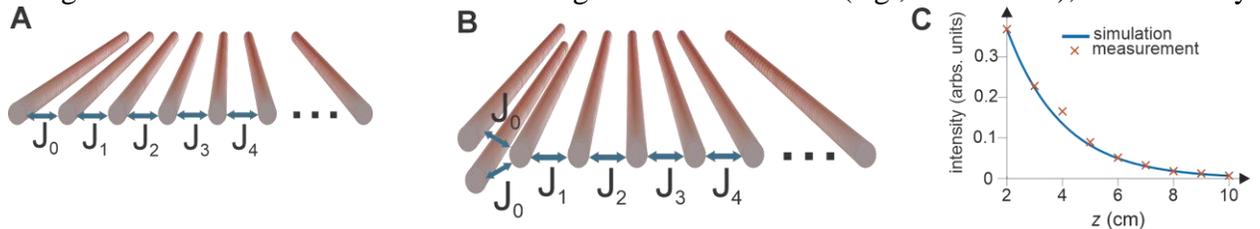

**Fig. S2. Lanczos transformation for one and two qubit systems**. Equivalent one-dimensional waveguide array system for (**A**) One qubit and (**B**) Two qubit systems, as illustrated in Figs. S1A and S1B, respectively. (**C**) Comparison of the measured intensity evolution in waveguide A across various sample lengths (red crosses) with the theoretical exponential decay characterized by a decay constant of $\Gamma = 0.25 \ cm^{-1}$ (solid blue line).

The following coupling coefficients has been used for both designs:

| n | | 0 | 1 | 2 | 3 | 4 | 5 | 6 | 7 | 8 | 9 | 10 | 11 | 12 | 13 | 14 | 15 | 16 |
|---|---|---|---|---|---|---|---|---|---|---|---|---|---|---|---|---|---|---|
| $J_n$ | | 0.798 | 2.309 | 2.066 | 2.028 | 2.016 | 2.010 | 2.007 | 2.005 | 2.004 | 2.003 | 2.003 | 2.002 | 2.002 | 2.001 | 2.001 | 2.001 | 2.001 |
| n | 17 | 18 | 19 | 20 | 21 | 22 | 23 | 24 | 25 | 26 | 27 | 28 | 29 | 30 | 31 | 32 | 33 | 34 |
| $J_n$ | 2.001 | 2.001 | 2.001 | 2.001 | 2.001 | 2.001 | 2.001 | 2.000 | 2.000 | 2.000 | 2.000 | 2.000 | 2.000 | 2.000 | 2.000 | 2.000 | 2.000 | 2.000 |
| n | 35 | 36 | 37 | 38 | 39 | 40 | 41 | 42 | 43 | 44 | 45 | 46 | 47 | 48 | 49 | 50 | 51 | |
| $J_n$ | 2.000 | 2.000 | 2.000 | 2.000 | 2.000 | 2.000 | 2.000 | 2.000 | 2.000 | 2.000 | 2.000 | 2.000 | 2.000 | 2.000 | 2.000 | 2.000 | | |

**Table S1. Coupling coefficients derived from the Lanczos transformation and utilized in the experiments.**

Notably, the length of the Lanczos array must be sufficiently large to prevent power coupled to the artificial environment from returning to the main waveguides (L, R). Furthermore, the number of waveguides in the Lanczos array determines whether the system is in the quasi-continuum limit. Figures S3A-C illustrates this concept, showing that as the number of waveguides in the environment increases, the power coupled to the environment eventually returns to the central waveguide (L, R) after a longer propagation distance.

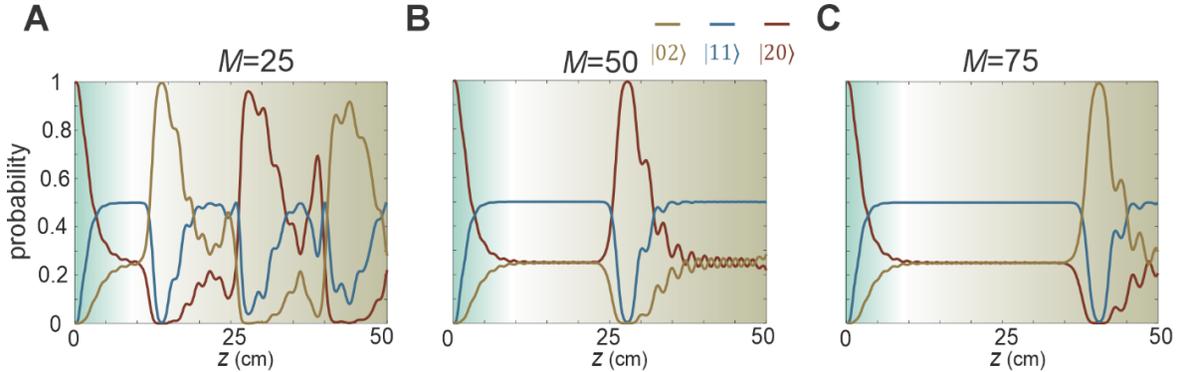

**Fig. S3. The effect of finite-length Lanczos array on the dynamics of the APT design.** Each plot shows the dynamics for different numbers of waveguides in the artificial environment: (**A**) $M=25$ (**B**) $M=50$ (**C**) $M=75$.

**Supplementary Text IV. Modeling anti-parity-time dynamics in a fully Hermitian way**
We examine the dynamics of a single photon in our system by applying the Heisenberg equation of motion, which governs the time evolution of the photon creation operators. This formalism allows us to track how the photon propagates and interacts within the framework of the system. The evolution equation could be written as:

$$i\frac{d}{dz}\begin{pmatrix}\hat{a}_L^\dagger\\\hat{a}_R^\dagger\\\hat{a}_1^\dagger\\\vdots\\\hat{a}_N^\dagger\end{pmatrix}=H\hat{\boldsymbol{a}}^\dagger=\begin{pmatrix}0 & 0 & J_L & 0 & .. & 0\\0 & 0 & J_R & J_1 & .. & 0\\J_L & J_R & \ddots & \ddots & & \vdots\\0 & J_1 & \ddots & & 0 & J_{N-1}\\\vdots & & & & & \\0 & 0 & 0 & J_{N-1} & & 0\end{pmatrix}\begin{pmatrix}\hat{a}_1^\dagger\\\vdots\\\hat{a}_N^\dagger\end{pmatrix}. \quad (S12)$$

The coupling strength $J_i$ between adjacent waveguides $i$ and $i-1$ depends on both the distance between the waveguides and the polarization of the photons. The coefficients $J_L$ and $J_R$ denote the coupling strengths for the left and right waveguides to the environment, respectively. For simplicity, we set $J_L = J_R = J_0$. The dynamics of photon propagation are governed by the transfer matrix $U = e^{iHz}$. The transmission probability for photons injected into the left or right waveguide is given by $|U_{i,i}(z)|^2$.

Using the coupling coefficients discussed previously, the steady-state eigenmode solution is $|\psi^{(1)}\rangle = \frac{1}{\sqrt{2}}(|10\rangle - |01\rangle)$. The probability of finding the system in the $N$-photon subspace—equivalent to detecting $N - n$ photons at port 1 and $n$ photons at port 2—is given by:

$$P_n^{(N)} = \frac{\left|\sum_{m=0}^N U_{n,m}^{(N)} c_m^{(N)}\right|^2}{\sum_{n=0}^N \left|\sum_{m=0}^N U_{n,m}^{(N)} c_m^{(N)}\right|^2}, \quad (S13)$$

where $U_{m,n}^{(N)}(z) = \langle N-m, m|\hat{U}(z)|N-n, n\rangle$ represents the matrix elements of the evolution operator $\hat{U}(z) = e^{iH^{(N)}z}$ in the $N$-photon subspace governed by the Hamiltonian $H^{(N)}$, with elements:

$$H_{m,n}^{(N)} := \langle N-m, m|\hat{H}|N-n, n\rangle = J_n \delta_{m,n-1} + J_m \delta_{m,n+1}. \quad (S14)$$

where $J_n = \sqrt{(N-n+1)n}$, this leads to the following eigenstates for two- and three-photon states:
- For two photons: $|\psi^{(2)}\rangle = (|20\rangle + |02\rangle)/2 - |1,1\rangle/\sqrt{2}$.
- For three photons: $|\psi^{(3)}\rangle = 0.35|30\rangle - 0.35|03\rangle + 0.61|2,1\rangle - 0.61|1,2\rangle$.
- For four photons: $|\psi^{(4)}\rangle = 0.25|40\rangle + 0.25|04\rangle - 0.5|3,1\rangle - 0.5|1,3\rangle + 0.61|2,2\rangle$.

Note that the three-photon and four-photon states can be measured using the proposed setup depicted in Fig. S4A and its variants. Interestingly, this behavior, characterized by a single attractor state in each photon-number subspace, starkly contrasts with other non-Hermitian structures. For example, in a parity-time (PT) symmetric coupler, the normalized probabilities oscillate periodically. In contrast, the APT filter exhibits fundamentally different dynamics, where the input states always equilibrate to the attractor of the APT system, as shown in Fig. S4B.

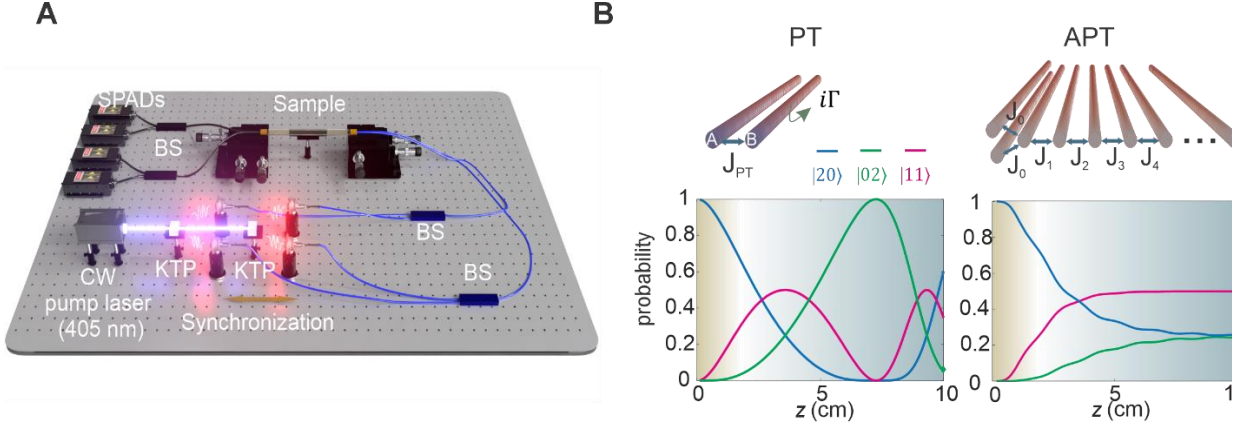

**Fig. S4. Higher photon number suggested measurement setup and comparison between PT and APT system evolution.** (**A**) Proposed experimental setup. Here, f-PBS: fiber polarizing beam splitter; BS: beam splitter; SPAD: single-photon avalanche diode. (**B**) Two photon dynamics for PT and APT structures. Here, we set $\Gamma = 0.25\ cm^{-1}$. In all cases, the probabilities are normalized.

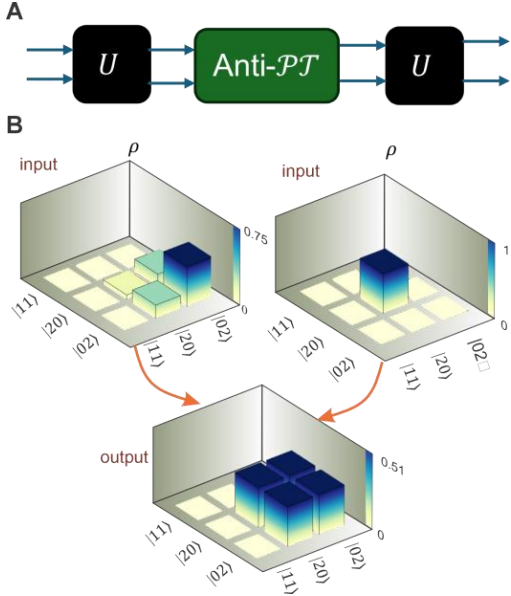

**Fig. S5. Arbitrary state filter**. (**A**) Schematic representation of a general state filter utilizing two successive unitary transformations. (**B**) A NOON state entanglement filter specifically designed to convert an arbitrary input state into the NOON state $|\psi\rangle = (|20\rangle + |02\rangle)/\sqrt{2}$.

**Supplementary Text V. Arbitrary pure state filter**

Thus far, our study has focused on the entanglement filter's capacity to isolate specific entangled states, such as $(|10\rangle + |01\rangle)/\sqrt{2}$ and $(|20\rangle + |02\rangle)/2 - |11\rangle/\sqrt{2}$. However, our filter is versatile and capable of outputting any pure state, whether entangled or not, by leveraging the techniques outlined in this section. In the single-photon subspace, any desired state can be generated by adjusting the coupling ratios of the $J_L$ and right $J_R$ qubits, as described in Supplementary Text I. Under these conditions, the steady state of the system, characterized by zero loss in the Hamiltonian [Eq. (S12)], is given by:

$$|\psi\rangle = \frac{1}{C}\left(-\frac{J_L}{J_R}|10\rangle + |01\rangle\right), \tag{S15}$$

where $C = \sqrt{\left(\frac{J_L}{J_R}\right)^2 + 1}$ represents the normalization constant. However, in the two-photon regime, states ranging from $(|20\rangle + |02\rangle)/2 - |11\rangle/\sqrt{2}$ (when $J_L = J_R$) to $|20\rangle$ (when $J_L \gg J_R$) can be produced. However, given the significant expansion of the Hilbert space for biphotons, not all states can be reached by simply varying the coupling ratios. Instead, the remaining states can be accessed through two unitary transformations, as demonstrated in Fig. S5A. For instance, applying the unitary transformation

$$U = \begin{pmatrix} 0.8556 & -0.5 & -0.1463 \\ 0.5i & 0.707i & 0.5i \\ -0.1463 & -0.5 & 0.8556 \end{pmatrix}$$

(where the computational basis is arranged such that $|\psi\rangle =$

$c_1|20\rangle + c_2|11\rangle + c_3|02\rangle \to |\psi\rangle = \begin{pmatrix} c_1|20\rangle \\ c_2|11\rangle \\ c_3|02\rangle \end{pmatrix}$ enables the conversion of our APT filter into a NOON state filter, as depicted in Fig. S5B. This transformation can be obtained through the combination of a nonlinear phase shift gate $U_1 = \begin{pmatrix} 1 & 0 & 0 \\ 0 & i & 0 \\ 0 & 0 & 1 \end{pmatrix}$ and a directional coupler or beam splitter with a two-photon transfer function $U_2 = \begin{pmatrix} 0.8556 & 0.5i & -0.1463 \\ 0.5i & 0.707 & 0.5i \\ -0.1463 & 0.5i & 0.8556 \end{pmatrix}$ (43). The transfer function $U_2$ itself can be derived from a beam splitter with the following single-photon transfer function: $U^{(1)} = \begin{pmatrix} 0.928 & 0.385i \\ 0.385i & 0.928 \end{pmatrix}$. The same outcome of selecting a NOON state can be achieved through post-selection on anti-coincidence events.

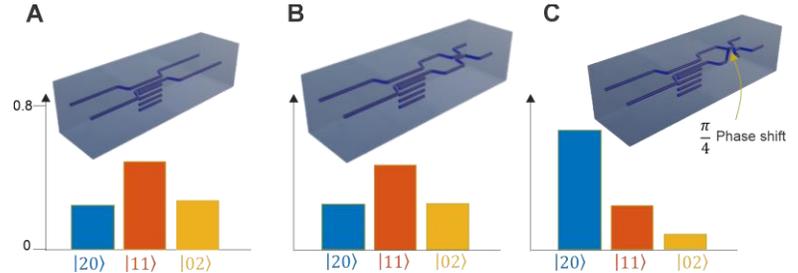

Fig. S6. The samples used for quantum state tomography measurements. (A-C) Measurement results for each configuration depicted in the inset. All anti-PT symmetric designs were implemented in 10 cm-long structures, employing a balanced 50:50 coupler for the experiments.

**Supplementary Text VI. Methods and sample fabrication**

The waveguide systems used in our experiments were fabricated with the femtosecond laser direct writing technique (39). To this end, ultrashort pulses from a frequency-doubled fibre amplifier system (Coherent MONACO) at a wavelength of 517 nm with a pulse length of 270 fs and a repetition rate of 333 kHz were focused into the volume of a 100 mm long fused silica sample (Corning 7980). The sample was translated by means of a precision positioning system (Aerotech ALS150) with respect to the focal spot, resulting in single-mode waveguides with a mode field diameter of approximately 13μm × 15μm for the probe wavelength of 810 nm. Propagation losses in the straight waveguides were measured to be below 0.3 dB/cm. The desired nearest-neighbour couplings ranging from approximately $0.8 \text{ cm}^{-1}$ to $2.3 \text{ cm}^{-1}$ were obtained by appropriately varying the separation between each pair of adjacent waveguides from 20.0 μm to 14.3 μm. To facilitate coupling to the fiber arrays, fanning sections were appended before and after the functional regions of the waveguide arrangements, increasing the separation between the channels that were to be excited and interrogated to 127 μm.

**Supplementary Text VII. Quantum state tomography measurements**

To elucidate the phase relationships between quantum states, we performed quantum state tomography measurements (44). We utilized the input state $|\psi_{in}\rangle = |11\rangle$ varying the phase measurement configurations across three distinct samples:
1. **Sample 1**: Contains only the anti-PT coupler.
2. **Sample 2**: Features the anti-PT coupler followed by a balanced coupler at its output.
3. **Sample 3**: Similar to Sample 2 but includes an additional π/4 phase shift before the balanced coupler.

These setups are illustrated in Figs. S5A-C. The balanced coupler is designed to induce self-interference of the quantum state. For example, the resulting probabilities for Sample 2 were $P_{20} = 0.2578, P_{02} = 0.2633$, and $P_{11} = 0.4789$—closely match the expected values for the anti-PT coupler output state $|\psi_{anti-\mathcal{PT}}\rangle = (|20\rangle + |02\rangle)/2 - |11\rangle/\sqrt{2}$.

However, the measured state, as shown in Fig. S6A, was $|\psi\rangle = 0.2438|20\rangle + 0.2677|02\rangle + 0.4885|11\rangle$, which deviates from the ideal state due to factors such as longer relaxation times for the two-photon state and experimental inaccuracies, including measurement and fabrication tolerances. It is noteworthy that the state $|\psi^{(2)}\rangle = (|20\rangle + |02\rangle)/2 + |11\rangle/\sqrt{2}$ can yield the same probabilities for the measurements shown in Fig. S6B.

To further differentiate between the quantum states, an additional measurement was conducted with a phase shift of $\phi = \pi/4$ introduced in one of the arms. This measurement yielded probabilities $P_{20} = 0.6737, P_{02} = 0.0816$, and $P_{11} = 0.2447$ (Fig. S6C). These results do not align with the probabilities predicted for the state $|\psi^{(2)}\rangle = (|20\rangle + |02\rangle)/2 + |11\rangle/\sqrt{2}$ and can only be reconciled with the state $|\psi^{(2)}\rangle \approx (|20\rangle + |02\rangle)/2 - |11\rangle/\sqrt{2}$.

## Supplementary Text VIII. Uniqueness of the phase determined by quantum state tomography measurements

The output state from the entanglement filter is:

$$|\psi_i\rangle = (|20\rangle + |02\rangle e^{i\phi_1})/2 - |11\rangle e^{i\phi_2}/\sqrt{2}. \quad (S16)$$

Let $a^\dagger$ and $b^\dagger$ represent the inputs to the balanced directional coupler. Thus, we can express the wavefunction at the coupler's input as:

$$|\psi_i\rangle = (a^{\dagger 2}|00\rangle + b^{\dagger 2}|00\rangle e^{i\phi_1})/2\sqrt{2} - a^\dagger b^\dagger |00\rangle e^{i\phi_2}/\sqrt{2}. \quad (S17)$$

Since we are using a balanced directional coupler, the output operators $c^\dagger$ and $d^\dagger$ are defined as $a^\dagger \to (c^\dagger + id^\dagger)/\sqrt{2}$, $b^\dagger \to (ic^\dagger + d^\dagger)/\sqrt{2}$. Or, in a more concise form:

$$|\psi\rangle = \left\{\left[\frac{1-e^{i\phi_1}}{4\sqrt{2}} - \frac{ie^{i\phi_2}}{2\sqrt{2}}\right]c^{\dagger 2} + \left[\frac{-1+e^{i\phi_1}}{4\sqrt{2}} - \frac{e^{i\phi_2}i}{2\sqrt{2}}\right]d^{\dagger 2} + \left[\frac{i(1+e^{i\phi_1})}{2\sqrt{2}}\right]c^\dagger d^\dagger\right\}|00\rangle. \quad (S18)$$

Then equating with $|\psi\rangle = c_1|20\rangle + c_2|02\rangle + c_3|11\rangle$, we obtain:

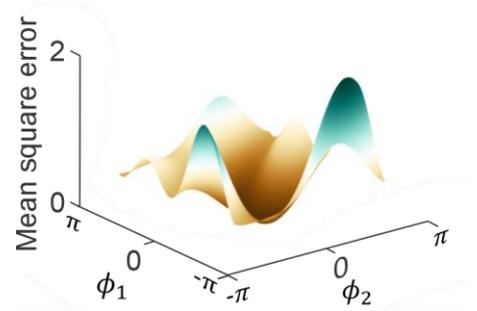

**Fig. S7. The mean square error for Equations (S24) and (S19) versus $\phi_1$ and $\phi_2$.** The presence of a global minimum where the mean square nulls indicate the uniqueness of the solution. The probabilities values $|c_1|^2 = 0.2578$, $|c_2|^2 = 0.4789$, $|c_3|^2 = 0.2633$, $|c_4|^2 = 0.6737$, $|c_5|^2 = 0.2447$, $|c_6|^2 = 0.0816$.

$$\left[\frac{1-e^{i\phi_1}}{4\sqrt{2}} - \frac{ie^{i\phi_2}}{2\sqrt{2}}\right]\left[\frac{1-e^{-i\phi_1}}{4\sqrt{2}} + \frac{ie^{-i\phi_2}}{2\sqrt{2}}\right] = |c_1|^2,$$

$$\left[\frac{-1+e^{i\phi_1}}{4\sqrt{2}} - \frac{e^{i\phi_2}i}{2\sqrt{2}}\right]\left[\frac{-1+e^{-i\phi_1}}{4\sqrt{2}} + \frac{e^{-i\phi_2}i}{2\sqrt{2}}\right] = |c_2|^2, \quad (S19)$$

$$\left[\frac{i(1+e^{i\phi_1})}{2\sqrt{2}}\right]\left[-\frac{i(1+e^{-i\phi_1})}{2\sqrt{2}}\right] = |c_3|^2.$$

The output state from the entanglement filter with a $\pi/4$ phase shift is given by:

$$|\psi_i\rangle = (|20\rangle + i|02\rangle)/2 - e^{\frac{i\pi}{4}}|11\rangle/\sqrt{2}. \quad (S20)$$

We can express the wavefunction at the input of the coupler as:
$$|\psi_i\rangle = (a^{\dagger 2}|00\rangle + e^{i\phi_1}e^{\frac{i\pi}{2}}b^{\dagger 2}|00\rangle)/2\sqrt{2} - e^{i\phi_2}e^{\frac{i\pi}{4}}a^{\dagger}b^{\dagger}|00\rangle/\sqrt{2}. \quad (S21)$$

At the output, the wavefunction becomes:
$$|\psi\rangle = ((c^{\dagger} + id^{\dagger})^2|00\rangle + e^{i\phi_1}e^{\frac{i\pi}{2}}(ic^{\dagger} + d^{\dagger})^2|00\rangle)/4\sqrt{2} - e^{\frac{i\pi}{4}}e^{i\phi_2}(c^{\dagger} + id^{\dagger})(ic^{\dagger} + d^{\dagger})|00\rangle/2\sqrt{2}. \quad (S22)$$

Simplifying further, we get:
$$|\psi\rangle = \left[c^{\dagger 2}\left(\frac{1 - e^{i\phi_1}e^{i\pi/2}}{4\sqrt{2}} - \frac{ie^{i\phi_2}e^{i\pi/4}}{2\sqrt{2}}\right) + d^{\dagger 2}\left(\frac{e^{i\phi_1}e^{i\pi/2} - 1}{4\sqrt{2}} - \frac{ie^{i\phi_2}e^{i\pi/4}}{2\sqrt{2}}\right) + ic^{\dagger}d^{\dagger}\left(\frac{e^{i\phi_1}e^{i\pi/2} - 1}{2\sqrt{2}}\right)\right]|00\rangle. \quad (S23)$$

This can be represented as superposition of basis states $|\psi\rangle = c_4|20\rangle + c_5|02\rangle + c_6|11\rangle$. Thus, we have the following set of equations:
$$\left(\frac{1 - e^{i\phi_1}e^{i\pi/2}}{4\sqrt{2}} - \frac{ie^{i\phi_2}e^{i\pi/4}}{2\sqrt{2}}\right)\left(\frac{1 - e^{-i\phi_1}e^{-i\pi/2}}{4\sqrt{2}} + \frac{ie^{-i\phi_2}e^{-i\pi/4}}{2\sqrt{2}}\right) = |c_4|^2,$$
$$\left(\frac{e^{i\phi_1}e^{i\pi/2} - 1}{4\sqrt{2}} - \frac{ie^{i\phi_2}e^{i\pi/4}}{2\sqrt{2}}\right)\left(\frac{e^{-i\phi_1}e^{-i\pi/2} - 1}{4\sqrt{2}} + \frac{ie^{-i\phi_2}e^{-i\pi/4}}{2\sqrt{2}}\right) = |c_5|^2, \quad (S24)$$
$$\left(\frac{e^{i\phi_1}e^{i\pi/2} - 1}{2\sqrt{2}}\right)\left(\frac{e^{-i\phi_1}e^{-i\pi/2} - 1}{2\sqrt{2}}\right) = |c_6|^2.$$

Using Eqs. (S24) and (S19), we have six equations in four unknowns (as $e^{i\phi_{1,2}} = X_{1,2} + iY_{1,2}$) which can be solved by minimizing the mean square error as $MSE = \sum_i^6 \left(|c_{i,exp}|^2 - |c_i(\phi_1, \phi_2)|^2\right)^2$, where $|c_{i,exp}|^2$ is the actual probabilities obtained from the experiments. Figure S7 shows the result of this process, indicating one global minimum which indicates the uniqueness of the solution.

**Supplementary Text IX. Hong-Ou-Mandel effect**
To synchronize the photons generated by the spontaneous parametric down-conversion (SPDC) source, we conducted a Hong-Ou-Mandel (HOM) experiment. The quantum source was coupled to a 50:50 fiber beam splitter, with a time delay introduced in one of the arms. The plot illustrates the delay-dependent coincidence counts (blue crosses), with error bars indicating the Poissonian standard deviation. A Gaussian fit (yellow line) yields a visibility of V=90.2% (Fig. S8), demonstrating a high degree of indistinguishability between the two photons.

**Supplementary Text X. Attractor state for N-photon**
Here, we derive the attractor state of the APT filter for the N-photon excitation. We begin by considering the following state:

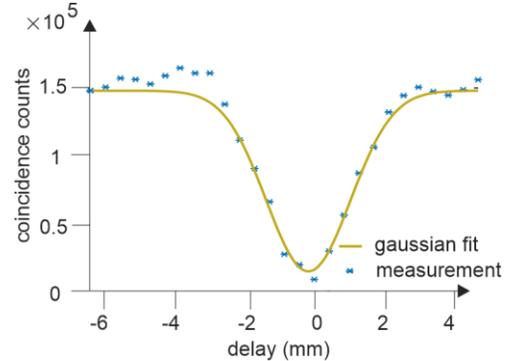

**Fig. S8. Characterization of photon pairs utilizing the Hong-Ou-Mandel (HOM) effect.** The quantum source is connected to a balanced fiber beam splitter, with a time delay introduced in one of the paths. The plot displays the coincidence counts as a function of the time delay (blue), with error bars indicating the Poissonian standard deviation. The fitted Gaussian curve (yellow) yields a visibility of V=90.2%, underscoring the high degree of indistinguishability between the two photons.

$$|\psi\rangle = c_0|N,0,0\rangle + c_1|N-1,0,1\rangle + c_2|N-2,0,2\rangle + \cdots + c_{N-1}|1,0,N-1\rangle \\ + c_N|0,0,N\rangle. \tag{S25}$$

The Hamiltonian of the system can be expressed as:
$$H_{int} = J_0 a_2^\dagger a_1 + J_0 \kappa a_2^\dagger a_3 + h.c.. \tag{S26}$$

Applying the Hamiltonian to the state $|\psi\rangle$, we obtain:
$$H_{int}|\psi\rangle = J_0\big(c_0\sqrt{N}|N-1,1,0\rangle + 0 + c_1|N-1,1,0\rangle + c_1\sqrt{N-1}|N-2,1,1\rangle \\ + c_2\sqrt{N-2}|N-3,1,2\rangle + c_2\sqrt{2}|N-2,1,1\rangle + \cdots\big) = 0. \tag{S27}$$

To have a steady state, we need to nullify this result. Thus, we require the following conditions:
$$c_0\sqrt{N} = -c_1 \to c_1 = -\frac{\sqrt{N}}{\sqrt{1}}c_0,$$
$$c_1\sqrt{N-1} = c_2\sqrt{2} \to c_2 = -\frac{\sqrt{N-1}}{\sqrt{2}}c_1, \tag{S28}$$
$$c_k = -\frac{\sqrt{N-(k-1)}}{\sqrt{k}}c_{k-1}.$$

Next, we apply the normalization condition. $c_0^2 + c_1^2 + c_2^2 + c_3^2 + \cdots + c_N^2 = 1$. This leads to:
$$c_0 = \sqrt{\frac{1}{\left(1 + N + \frac{N(N-1)}{2} + \frac{N(N-1)(N-2)}{2.3} + \cdots + 1\right)}} = \sqrt{\frac{1}{2^N}}. \tag{S29}$$

Moreover, the general form for $c_k = (-1)^k\sqrt{\frac{1}{2^N}\binom{N}{k}}$. As we can see, there is only one solution that makes the Hamiltonian vanish, indicating that our system has a unique steady state in each $N$-photon subspace. The same equations could be obtained using $\frac{1}{\mathbb{N}}\underbrace{|\psi_1\rangle \otimes \cdots \otimes |\psi_1\rangle}_{N-times} = \frac{1}{\mathbb{N}}\left(\frac{1}{\sqrt{2}}\right)^N\left(a_L^\dagger - a_R^\dagger\right)^N|00\rangle$ where $\mathbb{N}$ is a normalization constant.

### Supplementary Text XI. Decoherence-free subspace
The Hamiltonian of our system can be decomposed into three distinct parts:
$$H_{tot} = H_s + H_{int} + H_{env} \tag{S30}$$
Where $H_s = \delta\hat{a}_L^\dagger\hat{a}_L + \delta\hat{a}_R^\dagger\hat{a}_R$, $H_{int} = J_0(\hat{a}_L\hat{a}_1^\dagger + \hat{a}_R\hat{a}_1^\dagger + h.c.)$ and $H_{env} = \sum J_n\hat{a}_n\hat{a}_{n+1}^\dagger + h.c.$, here h.c. is the Hermitian conjugate. While $\delta$ is the detuning of each the left and right waveguide, which is taken to be zero for simplicity. The subspace is considered decoherence-free if the interaction Hamiltonian $H_{int}$ does not induce transitions between states within this subspace. Mathematically, this condition requires that for any state $H_{int}$ in the decoherence-free subspace $\mathcal{D}$, the expectation value of $H_{int}$ with respect to $|\psi\rangle$ must be zero, i.e., $\langle\psi|H_{int}|\psi\rangle = 0$ *(31)*.

Simultaneously, the subspace must be invariant under the action of $H_s$. This invariance means that the state $|\psi\rangle$ remains within the decoherence-free subspace $\mathcal{D}$ even as it evolves under $H_s$. This condition holds for the steady states in our system, whether for single photons, two photons, or more. This is further illustrated in Fig. S9A, where we excite single photon at a single site, while observing the system's response to randomly varying the coupling coefficient of the environment $(J_{1-50})$ by around 10%. Notably, the system still asymptotically reaches a steady state in all cases. Notably, once the system attains this steady state, it becomes locked in and exhibits immunity to

perturbations (Fig. S9B). Even after reaching this state, if the system is subjected to sudden perturbations in amplitude or phase, as shown in Figs. S9C-D, it self-heals, returning to a condition where the amplitudes of light in each waveguide are equal and the phase is approximately $\pi$.

**Supplementary Text XII. Mixed state purification**
The APT design, in a striking display of quantum filtering, transforms any mixed input state into a pristine pure state, anchoring it firmly within the steady state of the system. This robust invariance of the output state, regardless of input complexity, can be experimentally validated for single photons in Figs. 2C–D. Since as we will show in this section, such an output for symmetric system indicates that the structure capable of purifying mixed states. From these two figures we showed that whether input state $|10\rangle$ or $|01\rangle$ the output state will be $\approx \frac{1}{\sqrt{2}}(|10\rangle - |01\rangle)$. In other words, APT setup ensures that an input state like $|10\rangle$ or $|01\rangle$ yields $|\psi^{(1)}\rangle = U_{APT}|10\rangle = U_{APT}|01\rangle \approx \frac{k}{\sqrt{2}}(|10\rangle - |01\rangle)$, where $k = 1/\sqrt{2}$ represents the overlap with the steady state and $U_{APT}$ is the filter transfer function (which is not a unitary matrix).

Without loss of generality, we assume an arbitrary mixed input state, expressible as the superposition $|\psi_{input}\rangle = (c_{10}|10\rangle - c_{01}e^{i\theta(t)}|01\rangle)$, where $c_{10}, c_{01}$ are normalized constants such $|c_{10}|^2 + |c_{01}|^2 = 1$ and $\theta(t)$ is phase randomly varying with time. This wavefunction could be also recast in density matrix representation as $\hat{\rho} = \frac{1}{2}|10\rangle\langle 10| + \frac{1}{2}|01\rangle\langle 01|$. This state after APT propagation will end to $|\psi_{output}\rangle = U_{APT}|\psi_{input}\rangle = (c_{10}\frac{k}{\sqrt{2}}(|10\rangle - |01\rangle) - c_{01}e^{i\theta(t)}\frac{k}{\sqrt{2}}(|10\rangle -$

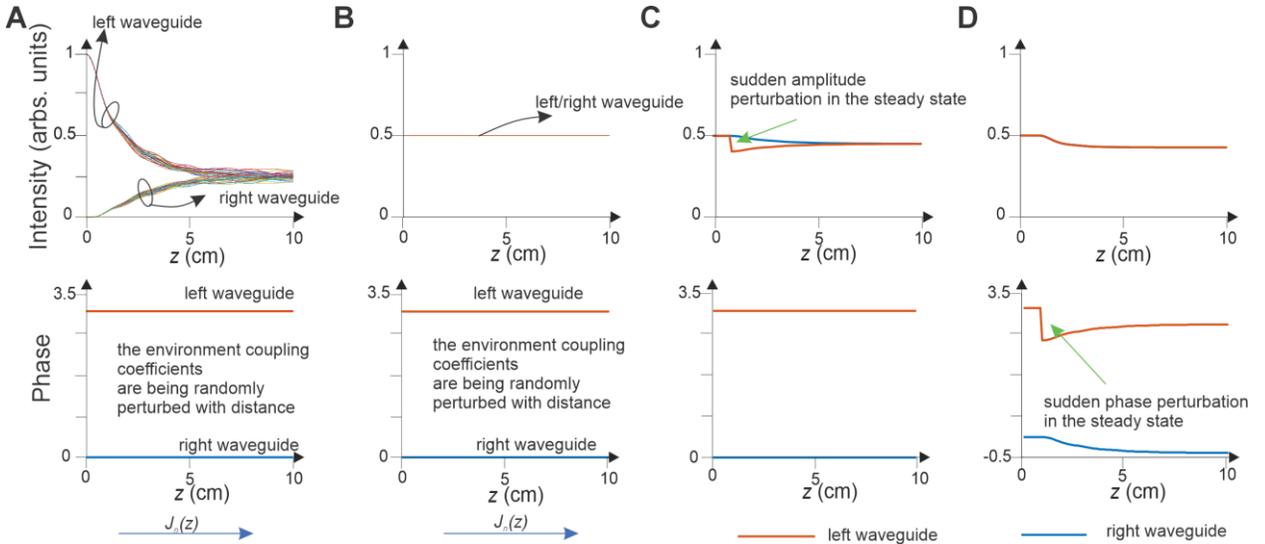

**Fig. S9. Effect of perturbations on the APT filter steady state**. The system's ensemble evolution is shown with environment coupling perturbed randomly along the z-axis by approximately 10% of the coupling coefficients under two different excitations. (**A**) Only the left waveguide is excited. (**B**) Both left and right waveguides are excited equally with a $\pi$ phase shift, representing the steady state of the APT filter. The influence of sudden perturbations on the steady state is depicted for (**C**) Amplitude and (**D**) Phase.

$|01\rangle)) = (|10\rangle - |01\rangle)(c_{10}\frac{k}{\sqrt{2}} - e^{i\theta(t)}c_{01}e^{\theta(t)}\frac{k}{\sqrt{2}}) = \frac{1}{\mathbb{N}}(|10\rangle - |01\rangle)$, with $\mathbb{N}$ is another normalization constant, regardless of the phase $\theta(t)$. With multi-photon interactions, this purification continues seamlessly. For two photons, our measurement basis $(|20\rangle, |02\rangle, |11\rangle)$ each aligns as $|\psi_{input}^{(2)}\rangle = |\psi_{in,L}^{(1)}\rangle \otimes |\psi_{in,R}^{(1)}\rangle$, here $\psi_{in,L(R)}^{(1)}$ denotes the single photon state for left (right) waveguide. Thus, $|\psi_{output}^{(2)}\rangle = \sum_i e^{i\theta_i(t)} U_{APT}|\psi_{in,L}^{(1)}\rangle \otimes U_{APT}|\psi_{in,R}^{(1)}\rangle \approx \frac{1}{2\sqrt{2}}\sum_i e^{i\theta_i(t)}(a_L^\dagger - a_R^\dagger)^2|00\rangle$. Similarly, a general two photon mixed state $|\psi_{input}^{(2)}\rangle = (c_{20}|20\rangle + c_{02}|02\rangle e^{i\theta_1(t)}) + c_{11}|11\rangle e^{i\theta_2(t)}$ will evolves to normalized wavefunction $|\psi_{output}^{(2)}\rangle = [(|20\rangle + |02\rangle)/2 - |11\rangle/\sqrt{2}]$.

To clarify this point, we measure the purity of the state during evolution through the trace of the square of the density matrix i.e., $Tr(\rho^2)$. Clearly, this quantity is unity for a pure state while for a completely mixed state is equal to $1/d$, where $d$ is dimension of the system (in our two-level APT configuration $d = 2$). Figure S10A illustrates the evolution of this 'purity' measure for a fully mixed state $\rho_{in} = \frac{1}{2}(|11\rangle\langle 11| + |20\rangle\langle 20|)$ when launched into the entanglement filter. After a propagation distance $z \gg 1/\Gamma$, the state becomes fully purified and settles into the entangled state $(|\psi\rangle = \frac{|11\rangle}{\sqrt{2}} - (|20\rangle + |02\rangle)/2$. It is noteworthy that the same dynamics ensue when $\rho_{in} = \frac{1}{2}(|11\rangle\langle 11| + |02\rangle\langle 02|)$ because of the system's symmetry. Similarly, if the input state is again fully mixed $\rho_{in} = \frac{1}{2}(|20\rangle\langle 20| + |02\rangle\langle 02|)$, it will undergo purification, albeit at a longer equilibration time (distance). This is because the state has a lower average overlap with APT arrangement than its counterpart $\rho_{in} = \frac{1}{2}(|11\rangle\langle 11| + |02\rangle\langle 02|)$, as shown in Fig. S10B. Moreover, as expected, the same dynamics persist even when the input state is partially mixed, as illustrated in Fig. S10C below. Next, we turn our attention to the entanglement dynamics. For this purpose, we utilize the participation function $\zeta$ as the entanglement measure, defined as $\xi = \frac{1}{\sum_{i=1}^k \lambda_i^2}$, where $\lambda_i$ is the eigenvalue of the reduced density matrix for either the left or right waveguide (45). This quantity is unity for a separable state and higher than unity for entangled state. As indicated in Fig. S10D, the separable state $|\psi\rangle = |11\rangle$ evolves into the entangled state $|\psi\rangle = \frac{1}{\sqrt{2}}|11\rangle - \frac{1}{2}(|20\rangle + |02\rangle)$, with the participation ratio increasing until it saturates when the system reaches steady state. Note that the same trends could be observed if we used entanglement measure as Renyi entropy defined as $S_\alpha(z) = \frac{1}{1-\alpha}\log_2 Tr[\rho_r^2(z)]$, where $\rho_r$ is the reduced density matrix for either the left or right waveguide, and $\alpha$ is a free parameter, which we set to 2 for simplicity.

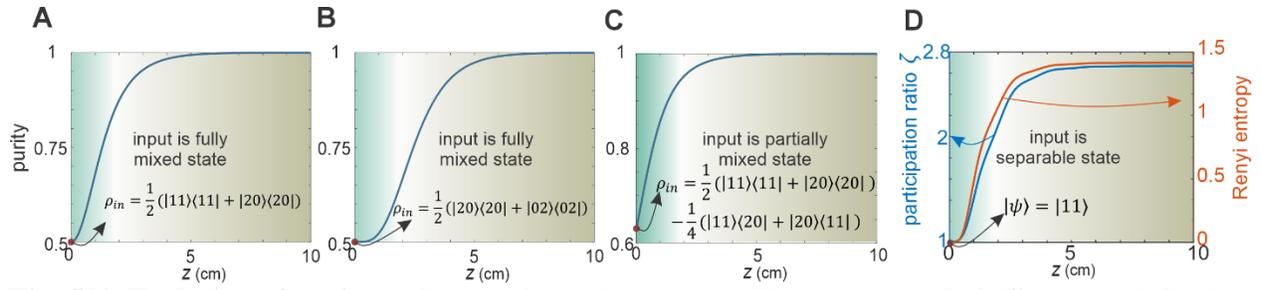

**Fig. S10. Evolution of purity and entanglement measures**. The dynamics of (**A-C**) purity (defined as $Tr(\rho^2)$) for various input states and (**D**) Participation ratio and Renyi entropy dynamics for input state $|\psi\rangle = |11\rangle$.